\documentclass[aps,twocolumn,gbroupedaddress,amsmath,amssymb]{revtex4-2}
\usepackage[utf8]{inputenc}
\usepackage[T1]{fontenc}
\usepackage{amsmath}
\usepackage{amsfonts}
\usepackage{amssymb}
\usepackage{graphics,graphicx}

\usepackage{graphicx}
\usepackage{subfigure} %colocar figura lado a lado, figura a) e b).
\usepackage{wrapfig} % pacote reponsavel para colocar figura ao lado do texto
\usepackage{epstopdf}
\usepackage{xcolor}
\graphicspath{{figuras/}}

\usepackage{ulem}
\usepackage[breaklinks=true]{hyperref}
\usepackage{setspace}
\usepackage{graphicx}
\usepackage{color}

\begin{document}

\title{Exploring the shear viscosity in four-dimensional planar black holes beyond General Relativity}

\author{Mois\'es Bravo-Gaete}
\email{mbravo-at-ucm.cl} \affiliation{Departamento de Matem\'aticas, F\'isica y Estad\'istica, Facultad de Ciencias
B\'asicas, Universidad Cat\'olica del Maule, Casilla 617, Talca,
Chile.}

\author{Luis Guajardo}
\email{luis.guajardo.r-at-gmail.com} \affiliation{Instituto de Investigaci\'on Interdisciplinaria, Vicerrector\'ia Acad\'emica, Universidad de Talca, Talca -- Chile.}

\author{Fabiano F. Santos}
\email{fabiano.ffs23-at-gmail.com} \affiliation{Instituto de Física,Universidade Federal do Rio de Janeiro, Caixa Postal 68528, Rio de Janeiro-RJ, 21941-972 -- Brazil}

\begin{abstract}
The well known shear viscosity to entropy density ratio ($\eta /s$) cannot be computed when the black hole space-time has zero thermodynamic entropy. This is the case, for example, when General Relativity in four dimensions is complemented with Critical Gravity, or in particular scenarios within the \textit{Four-dimensional-scalar-Gauss-Bonnet} theories. Recently, it has been shown that the zero entropy situation can be overcome in these examples by introducing suitable matter fields. With this at hand, in this paper we analyze each case and the impact of these extra sources in the ratio, in terms of their new parameters. We find that, while the $\eta /s$ ratio remains constant and insensitive in the former, this is not the case for the latter. To perform the calculations, we construct a Noether charge using a space-like Killing vector. The accuracy of the aforementioned findings is supported by the Kubo formalism.
\end{abstract}

\maketitle
%\tableofcontents
\newpage

%%%%%%%%%%%%%%%%%%%%%%%
\section{Introduction}
%%%%%%%%%%%%%%%%%%%%%%%

The image of the Black Holes (BH) located at the center of galaxies M87 and the Milky Way, captured by the Event Horizon Telescope \cite{EventHorizonTelescope:2019dse}-\cite{EventHorizonTelescope:2022xqj}, represents one of the most intriguing predictions of General Relativity (GR). Their essential feature is that they can be studied as thermodynamic systems. In fact, and according to a series of papers from Bardeen, Bekenstein, Carter and Hawking (see \cite{Hawking:1971vc}-\cite{Hawking:1975vcx}), they satisfy the so-called Four laws of Black Hole Thermodynamics, relating physical quantities, such as the temperature (Hawking Temperature) and the entropy (Bekenstein-Hawking entropy) of the system, to the surface gravity and the area of its event horizon, respectively. {In particular, Hawking showed using a semi-classical approach, i.e., when quantum mechanical effects near the horizon are taken into account, that a black hole will emit particles as if were a hot body \cite{Hawking:1971vc}. The Hawking Temperature, and the Bekenstein-Hawking entropy are remarkable results that points towards a link between Gravity and Quantum Mechanics.
}

{In this context}, the Anti-de-Sitter/Conformal Field Theory (AdS/CFT) correspondence \cite{Maldacena:1997re}-\cite{Gubser:1998bc}, gauge/gravity duality, or simply holography, has taken the lead for the last twenty years. It conjectures a relation between strongly coupled quantum field theories and gravity theories in extra dimensions, in such a way that the physical behavior of a quantum system can be interpreted from its gravity dual. Within this framework, the study of field theories at finite temperatures naturally leads to the analysis of BHs, because from the gravity side of the duality they possess a well-defined temperature and enjoy rich thermodynamic properties.

Concretely, planar base manifolds are excellent candidates for studying quantum theories related to fluids. In fact, in translation invariance theories, the ideas from the duality permit us to extend thermodynamics into hydrodynamics, allowing the interpretation of  transport coefficients analyzing BH space-times, as was shown in \cite{Kovtun:2004de},  obtaining one of the most celebrated results of holography: the {ratio between shear viscosity ($\eta$) and density entropy ($s$)} (denoted by $\eta/s$) which has been postulated to satisfy a universal lower bound known as Kovtun-Son-Starinets (KSS)-bound,

\begin{eqnarray}\label{KSS}
\frac{\eta}{s} \geq \frac{1}{4 \pi},
\end{eqnarray}
relating the event horizon to an ideal fluid \cite{Kovtun:2004de}-\cite{Kovtun:2003wp}. This conjecture has been proved for a large list of gravity theories and received some experimental support, for example  a superfluid Fermi gas \cite{Rupak:2007vp} and, in quantum chromodynamics, the quark-gluon plasma \cite{Song:2010mg,Shuryak:2003xe}.
Nevertheless, it is possible to obtain some concrete examples where its universality has been a subject of debate (see for example \cite{Brigante:2007nu}-\cite{Bravo-Gaete:2020lzs}). Some of them are characterized by the presence of higher-order curvature gravity or the inclusion of suitable matter sources, as well as some anisotropic theories. Also, the bound (\ref{KSS}) can be violated in massive gravity theories
\cite{Baggioli:2016rdj}. According  \cite{Alberte:2016xja}, the KSS bound may be breached in solid materials with a non-zero elastic modulus that exhibit viscoelastic mechanical responses.

In this work, we are motivated by concrete examples beyond GR in which the black hole space-time features a vanishing thermodynamic entropy, making the ratio $\eta/s$ ill-defined. In these examples, the inclusion of extra matter fields provides a route to circumvent the zero entropy scenario in a proper way, making now an interesting question whether the new parameters of the theory affects the $\eta/s$ ratio (\ref{KSS}) or not.

First, we will analyze the scenario from Ref. \cite{Lu:2011zk} (hereafter, Case A), which includes the Einstein-Hilbert (EH) \textcolor{black}{Lagrangian} together with a cosmological constant
\begin{eqnarray}\label{eq:action-EHLambda}
\mathcal{L}_{\tiny{EH}}=\frac{1}{2} \left(R-2\Lambda\right),
\end{eqnarray}
supplemented with the so-called four-dimensional Critical Gravity (CG):
\begin{eqnarray}\label{eq:LCG}
\mathcal{L}_{\tiny{CG}}=\frac{1}{2}\left[
-\left(\frac{1}{2\Lambda}\right)\,{R}^2+\left(\frac{3}{2\Lambda}\right){R}_{\mu \nu}{R}^{\mu \nu}\right],
\end{eqnarray}
which enjoys to be a ghost-free, renormalizable theory of gravity with quadratic corrections in the curvature, despite the fact that the equations of motions are fourth-order. In Ref. \cite{Lu:2011zk}, a BH solution is obtained with null thermodynamic quantities. Nevertheless, as was shown in  \cite{Alvarez:2022upr},  { this situation} can be overcome through the introduction of a suitable matter source using nonlinear electrodynamics in the  Pleb\'anski formalism \cite{Plebanski:1968}, being the first example in four-dimensional CG where the BH thermodynamic parameters do not vanish. In this case, we will show below that the theory is insensitive to the parameter space, and the bound (\ref{KSS}) is always saturated.

On the other hand (Case B), in Ref. \cite{Correa:2013bza} the authors showed that it is possible to obtain exact higher-dimensional black hole solutions in a model given by Lanczos-Lovelock gravity theories imposing a unique AdS vacuum \cite{Crisostomo:2000bb}, dressed with a non-minimally coupled scalar field $\psi$. In dimensions greater than or equal to five, the simplest extension of Lanczos-Lovelock gravity is the Einstein-Gauss-Bonnet action coupled to a cosmological constant
\begin{equation}\label{EHGB}
S_{\tiny{EGB}}=\frac{1}{2}\int d^{D}x \sqrt{-g}\left[R-2\Lambda +\alpha \mathcal{L}_{GB}\right],
\end{equation} 
where $\mathcal{L}_{GB}=R^{2}-4R_{\mu \nu} R^{\mu \nu}+R_{\mu \nu \sigma \rho} R^{\mu \nu \sigma \rho}$ is the Gauss-Bonnet density. As it was done in \cite{Correa:2013bza}, at the moment to perform the thermodynamic study, the mass as well as the entropy vanish trivially and concluding that the integration constant of the solution can be interpreted as a sort of hair. The above issue was improved in \cite{Bravo-Gaete:2021hza} via the inclusion of a non-linear Maxwell source coupled to the scalar field $\psi$, making it possible to obtain a nontrivial thermodynamics analysis, in particular for the entropy, performing a natural exploration to obtain the shear viscosity $\eta$ and the analysis of (\ref{KSS}).  Inspired by the approach of previous work, and aiming to incorporate the Gauss-Bonnet density as an active contributor in lower dimensions,  the setup a four-dimensional background for the solution from Ref. \cite{Bravo-Gaete:2021hza} can be done through the recently proposed \textit{Four-dimensional-scalar-Gauss-Bonnet} (4DSGB) (see, e.g., \cite{Fernandes:2020nbq,Hennigar:2020lsl,Hennigar:2020fkv,Lu:2020iav,Bravo-Gaete:2022mnr,Kobayashi:2020wqy} and \cite{Fernandes:2022zrq} for a review) theory:
\begin{eqnarray}
\textcolor{black}{\mathcal{L}_{4DSGB}}&=&\frac{\tilde{\alpha}}{2}\Big(\phi \mathcal{L}_{GB}+4 G^{\mu \nu} \nabla_{\mu} \phi \nabla_{\nu} \phi \nonumber\\
&-&4 X \Box \phi+ 2 X^{2}\Big).\label{eq:GB-reg}
\end{eqnarray}
In the previous equation, $G_{\mu \nu}$ is the Einstein-Tensor, $\Box \phi= \nabla^{\mu} \nabla_{\mu} \phi$, $X=\nabla^{\sigma} \phi \nabla_{\sigma} \phi$ is the kinetic term and $\tilde{\alpha}$ is a constant obtained after a rescaling. Supplementing the Einstein-Hilbert \textcolor{black}{Lagrangian} (\ref{eq:action-EHLambda}) with eq. (\ref{eq:GB-reg}) introduces an additional scalar degree of freedom as the price-to-pay for reducing the Gauss-Bonnet model  to a four-dimensional background. In contrast to Case A, we will demonstrate in the following lines that some parameters have an influence on the $\eta/s$ ratio. Therefore, it is imperative to analyze the complete parameter space and determine how the $\eta/s$ ratio (\ref{KSS}) is affected in this scenario.

Traditionally, the shear viscosity $\eta$ can be calculated by effective coupling constants of the transverse
graviton on the location of the event horizon via the
membrane paradigm \cite{Iqbal:2008by} and the Kubo formula \cite{Cai:2008ph,Cai:2009zv}. Nevertheless, as we will show below, a recent formalism constructed via a Noether charge and a suitable election of a space-like Killing vector was proposed in \cite{Fan:2018qnt}, greatly simplifying the computations in comparison with the traditional techniques. This formulation is explained in detail in Section \ref{Section-gen-comp-shear}. With the above information,  the rest of the paper goes as follows: in Section \ref{Section-shear-sol}, we will explore the shear viscosity $\eta$ from four-dimensional planar BH's where the gravity theories are given by {the cases A-B}. Together with the above, we will analyze first how the $\eta/s$ ratio is affected under these theories beyond GR, and study in which cases the KSS-bound (\ref{KSS}) can be satisfied or violated. Additionally, the Appendix provides an explanation of how the $\eta/s$ ratio is re-derived from the Kubo formula. These results further confirm the findings presented in Section \ref{Section-shear-sol}. Finally, Section \ref{Section-conclusions} is devoted to our conclusions and discussions.

%%%%%%%%%%%%%%%%%%%%%%%%%%%%%%%%%%%%%%%%%%%%%%%%%%%%%
\section{Calculating the shear viscosity through a conserved charge} \label{Section-gen-comp-shear}
%%%%%%%%%%%%%%%%%%%%%%%%%%%%%%%%%%%%%%%%%%%%%%%%%%%%%

To be self-contained, in this section we will explain the construction that allows to obtain the shear viscosity $\eta$ following the procedure performed in \cite{Fan:2018qnt}, relating the Noether charge to a space-like Killing vector. As a first step, we vary the action
\begin{equation}\label{eq:gen-action}
S = \int d^4{x}\sqrt{-g} \Big( \mathcal{L}_{\tiny{g}}+\mathcal{L}_{\tiny{matter}} \Big)
\end{equation}
with respect to all the dynamical fields present in the theory, where $\mathcal{L}_{\tiny{g}}$ and $\mathcal{L}_{\tiny{matter}}$ represent the gravity and matter Lagrangians respectively, gives:
\begin{eqnarray}
\delta S&=&\delta \Big(\sqrt{-g} ( \mathcal{L}_{\tiny{g}}+\mathcal{L}_{\tiny{matter}}) \Big)\nonumber\\
&=&\sqrt{-g} \big[{\cal{E}}^{\mu \nu} \delta A_{\mu \nu} + {\cal{B}}^{\mu} \delta B_{\mu} + {\cal{G}}_{\varphi} \delta \varphi \nonumber\\
&+&\nabla_{\mu} J^{\mu} (\delta A, \delta B, \delta \phi)\big],
\end{eqnarray}
where ${\cal{E}}^{\mu \nu}$, ${\cal{B}}^{\mu}$ and ${\cal{G}}_{\varphi}$ denote collectively the equations of motion with respect to the tensors $A_{\mu \nu}$ (including the metric $g_{\mu \nu}$),  \textcolor{black}{the vectors} $B_{\mu}$ and the scalar fields $\varphi$ respectively, while  ${J}^{\mu}$ represents the surface term. \textcolor{black}{Starting from the current density $J^{\mu}$, we define a $1$-form $J_{(1)}=J_{\mu} dx^{\mu}$ and its Hodge dual $\Theta_{(3)}=(-1) * J_{(1)}$. Considering the variation induced by a diffeomorphism generated by a Killing vector $\xi_{\mu}$, which acts on the fields $A_{\mu\nu}$, $B_{\mu}$, and $\varphi$ according to 
\begin{eqnarray*}
\delta_{\xi} A_{\mu \nu}&=&\xi^{\sigma} \nabla_{\sigma} A_{\mu \nu}+(\nabla_{\mu} \xi^{\sigma}) A_{\sigma \nu}+(\nabla_{\nu} \xi^{\sigma}) A_{\sigma \mu},\\
\delta_{\xi} B_{\mu}&=&\xi^{\sigma} \nabla_{\sigma} B_{\mu}+(\nabla_{\mu} \xi^{\sigma}) A_{\sigma},\\
\delta_{\xi} \varphi&=&\xi^{\sigma} \nabla_{\sigma} \varphi,
\end{eqnarray*} 
where, when $A_{\mu \nu}=g_{\mu \nu}$, we recover $\delta_{\xi} g_{\mu \nu}=2 \nabla_{(\mu} \xi_{\nu)}$, and together with the equations of motions  ${\cal{E}}^{\mu \nu}=0$, ${\cal{B}}^{\mu}=0$ and ${\cal{G}}_{\varphi}=0$, we have that 
} 
$$J_{(3)}=\Theta_{(3)}-i_{\xi} * ( \mathcal{L}_{\tiny{g}}+\mathcal{L}_{\tiny{matter}}) =d(*J_{(2)}).$$
Here, $i_{\xi}$ represents a contraction of the vector $\xi^{\mu}$ with the first index of $* ( \mathcal{L}_{\tiny{g}}+\mathcal{L}_{\tiny{matter}})$, while that in our notations the sub-index "($p$)" represents the fact that we are working with $p$-forms.  With all the above, we can to define a $2$-form $Q_{2}=*J_{(2)}$ such that $J_{(3)}=d Q_{(2)}$, where $Q_{(2)}=Q_{\alpha_1 \alpha_2}=\epsilon _{\alpha_1 \alpha_2 \mu \nu} Q^{\mu \nu}$. \textcolor{black}{Here, $Q^{\mu \nu}$ is an antisymmetric tensor satisfies
\begin{eqnarray}\label{eq:Noether}
\nabla_{\nu} \mathcal{Q}^{\mu \nu}&=&   J^{\mu} (\delta_{\xi} A, \delta_{\xi} B, \delta_{\xi} \phi)-( \mathcal{L}_{\tiny{g}}+\mathcal{L}_{\tiny{matter}}) \xi^{\mu}.
\end{eqnarray} 
Now,} the next step is to perform a transverse and traceless perturbation on the four-dimensional line element
\begin{eqnarray}\label{eq:t-metric}
	ds^2&=&-h(r) dt^2+\frac{dr^2}{f(r)}+2 r^2 \Psi(t,r) dx dy+r^2dx^2\nonumber\\
&+&r^2dy^2,
\end{eqnarray}
where $t \in \mathbb{R}$, $r>0$, while we assume  $0\leq x \leq \sigma_{x}$ and $0\leq y \leq \sigma_{y}$. According to \cite{Fan:2018qnt}, for a space-like Killing vector $\partial_{x}= \xi^{\mu} \partial_{\mu}$ and with a linear time-dependent ansatz
\begin{equation}\label{eq:Psi}
\Psi(t,r) =\varsigma t +h_{xy}(r),
\end{equation}
where $\varsigma$  is a constant identified as the gradient of the fluid velocity along the $x$ direction, and at the transverse direction $y$ we can find from {eq. (\ref{eq:Noether}):}
\begin{eqnarray}
\nabla_{r} \mathcal{Q}^{ r y}=0,
\end{eqnarray} 
and the charge $\sqrt{-g}\mathcal{Q}^{ r y}$ becomes an integration constant \textcolor{black}{of the $(x,y)$-component of the linearized Einstein equations, determining the dynamics of the transverse perturbation}, allowing us to compute it from the near horizon solutions, and corresponding to the resistance of the shearing flows, this is, 
$$\sqrt{-g}\mathcal{Q}^{ r y}=\varsigma \eta,$$
where $\eta$ is the shear viscosity, which it can be obtained in the following way
\begin{eqnarray}\label{eq:shear}
	\eta=\frac{\partial \left(\sqrt{-g}\mathcal{Q}^{ r y}\right)}{\partial \varsigma}.
\end{eqnarray}
With all these ingredients, we are in condition to calculate the shear viscosity $\eta$, following this procedure, for \textcolor{black}{two} concrete examples in the following section.

%%%%%%%%%%%%%%%%%%%%%%%%%%%%%%%%%%%%%%
\section{Exploring the four-dimensional viscosity/entropy density ratio and analyzing the Kovtun-Son-Starinets bound} \label{Section-shear-sol}
%%%%%%%%%%%%%%%%%%%%%%%%%%%%%%%%%%%%%%

%%%%%%%%%%%%%%%%%%%%%%%%%%%%%%%%%%%%%%
{\subsection{Case A}}
%%%%%%%%%%%%%%%%%%%%%%%%%%%%%%%%%%%%%%

In this section, we will consider an electrically charged Anti-de Sitter black hole configuration, with the Lagrangian $\mathcal{L}_{\tiny{g}}$ from (\ref{eq:gen-action}) given by (\ref{eq:action-EHLambda})-(\ref{eq:LCG}) , and a matter source given by nonlinear electrodynamics in the Pleb\'anski formalism, represented through 
\begin{eqnarray}\label{eq:NLE}
\mathcal{L}_{\tiny{matter}}=-\frac{1}{2}F_{\mu\nu}P^{\mu\nu} + \mathcal{H}(P).
\end{eqnarray}
Here, $P^{\mu\nu}$ is an antisymmetric tensor conjugate to the field strength tensor $F_{\mu\nu}$, 
\begin{equation}\label{eq:Fmunu}
F_{\mu \nu}:=2\partial_{[\mu} A_{\nu]},
\end{equation}
while that $P = \dfrac{1}{4}P_{\mu\nu}P^{\mu\nu}$ is a scalar, and $\mathcal{H}(P)$ is an structural function depending on $P$.

As it was shown in \cite{Alvarez:2022upr}, in the line element (\ref{eq:t-metric}) with $\Psi(t,r)=0$ an electrically charged BH solution can be found, with
\begin{eqnarray}\label{eq:f}
h(r)=f(r) =r^2 \left(1-{\frac {\alpha_1\sqrt{M}}{r}}+{\frac {\alpha_2 M}{r^{2}}}-{\frac {\alpha_3 M^{3/2}}{r^{3}}}\right),
\end{eqnarray}
where $M$ is an integration constant, the $\alpha_{i}$'s correspond to structural coupling constants, while the structural function $\mathcal{H}(P)$, the electromagnetic field strength, as well as the cosmological constant $\Lambda$ take the form
\begin{eqnarray}
\mathcal{H}(P) &=&  \frac{1}{3} (\alpha_2^2-3 \alpha_1 \alpha_3)P-2\alpha_1 (-2 P)^{1/4}\\
&+&\alpha_2\sqrt{-2 P},\nonumber \\
F_{\mu\nu}&=& 2\delta^{t}_{[\mu}\delta^{r}_{\nu]} \left(\frac {r\alpha_1}{\sqrt {M}}-{
\alpha_2}-\frac{M \left(
3\,\alpha_1\,\alpha_3-\alpha_2^2 \right) }{3 r^2}\right), \nonumber\\
\Lambda&=&-3.\nonumber
\end{eqnarray}

Concerning their thermodynamic quantities, the electric charge and electric potential are given by 
\begin{eqnarray}
\mathcal{Q}_{e}&=&\frac{\Omega_{2} r_h^{2}}{\zeta^{2}},\\
\Phi_{e}&=&{r_h} \Big({\alpha_2}+
\alpha_1^{2}-\frac{3}{2} \alpha_1\zeta
-{\frac {\alpha_1\,\alpha_2}{\zeta}}+\frac{1}{3}\,
{\frac {\alpha_2^{2}}
{\zeta^{2}}}
\Big),
\end{eqnarray}
while that the mass, temperature and entropy read
\begin{eqnarray}
\mathcal{M}&=&
\frac{\alpha_1 \alpha_2 r_h^{3}\Omega_{2}}{9 \zeta ^{3}},\\
T&=&\frac{r_h}{4 \pi } \left(3-\frac{2 \alpha_1}{\zeta}+\frac{\alpha_2}{\zeta^{2}}\right),\label{eq:T-1}\\
\mathcal{S}&=&{2  \pi \Omega_{2}} r_h^{2}\left(\frac{\alpha_1}{\zeta}-\frac{2 \alpha_2}{3 \zeta^{2}}\right).\label{eq:ent1}
\end{eqnarray}
Here, $r_h$ represents the location of the event (or outer) horizon, which can be cast as $r_h=\zeta \sqrt{M} $, where $\zeta$ is a root of the cubic polynomial
$$\zeta^{3}-\alpha_1 \zeta^{2}+\alpha_2 \zeta-\alpha_3=0.\label{eq:entropy}$$
In the course of this work, $\Omega_{2}=\int dxdy=\sigma_{x} \sigma_{y}$ is the finite volume of the planar base manifold. Following the steps from \cite{Fan:2018qnt}, the entropy density $s$ reads
\begin{equation}\label{eq:s-1}
s=\frac{\mathcal{S}}{\Omega_{2}}={2 \pi} r_h^{2}\left(\frac{3 \zeta \alpha_1-2 \alpha_2}{3 \zeta^{2}}\right),
\end{equation}
and with the transverse and traceless perturbation (\ref{eq:t-metric}), as well as (\ref{eq:Psi}) with a space-like Killing vector $\partial_{x}= \xi^{\mu} \partial_{\mu}$, the $(x,y)$-component of the linear Einstein equations yields
\begin{eqnarray}
&&\Big\{-\frac{1}{4}f \Big[f r^2 (h_{xy})'''+2r(f+r f') (h_{xy})''\nonumber\\
&&+\frac{2}{3} (h_{xy})'(-f+4rf'+r^2f'''-3r^2)\Big]\Big\}'=0, 
\end{eqnarray}
where $(')$ denotes the derivative with respect to the radial coordinate $r$, implying that the charge $\sqrt{-g}\mathcal{Q}^{ r y}$ becomes an integration constant, where
\begin{eqnarray}\label{eq:ch-1-gen}
\sqrt{-g}\mathcal{Q}^{ r y}&=&-\frac{1}{4}f \Big[f r^2 (h_{xy})'''+2r(f+r f') (h_{xy})''\nonumber\\
&+&\frac{2}{3} (h_{xy})'(-f+4rf'+r^2f'''-3r^2)\Big].
\end{eqnarray}
Imposing the ingoing horizon boundary condition for $h_{xy}$ and a Taylor expansion for $h=f$ around the location of the event horizon $r_ h$:
\begin{equation*}
h_{xy}=\varsigma  \frac{\log(r-r_h)}{4 \pi T}+\cdots,\qquad h=f=4 \pi T (r-r_h)+\cdots,
\end{equation*}
where $T$ is the Hawking temperature (\ref{eq:T-1}), the charge (\ref{eq:ch-1-gen}) reads
\begin{eqnarray}\label{eq:ch-1}
\sqrt{-g}\mathcal{Q}^{ r y}&=&\frac{\varsigma  s }{4\pi},
\end{eqnarray}
allowing us to obtain the shear viscosity $\eta$ in the following way
\begin{eqnarray}\label{eq:eta-1}
&&\eta=\frac{s}{4\pi} \Rightarrow \frac{\eta}{s}=\frac{1}{4\pi},
\end{eqnarray}

{In} this situation, it is worth noting that the $\eta/s$ ratio remains constant, given by {the} saturated situation from (\ref{KSS}), and does not depend on any of the additional parameters of the theory, {namely, the coupling constants from CG} and the inclusion of non-linear electrodynamics. This behavior is analogous to the observed characteristics in GR and other situtations as shown in \cite{Kovtun:2004de}-\cite{Shuryak:2003xe}. {However,  the KSS bound (\ref{KSS}) can be affected by the parameters of the theory (see Refs. \cite{Brigante:2007nu}-\cite{Bravo-Gaete:2020lzs}). Indeed, this is what will occur in the following case.}
%%%%%%%%%%%%%%%%%%%%%%%%%%%%%%%%%%%%%%
{\subsection{Case B}
%%%%%%%%%%%%%%%%%%%%%%%%%%%%%%%%%%%%%%

As it was presented in the introduction, the five-dimensional action that was studied in ref. \cite{Bravo-Gaete:2021hza} is
\begin{equation}
S[g_{\mu\nu}, \psi, A_{\mu}] = S_{EGB} + S_{\psi} + S_{M},
\end{equation} with $S_{EGB}$ given by (\ref{EHGB}), and the matter sources are a {non-minimally coupled scalar field $\psi$,}
\begin{eqnarray}\label{eq:psi-action}
S_{\psi} =\int {d}^5 x\sqrt{-g} \mathcal{L}_{\psi}\nonumber\\
=\int {d}^5 x\sqrt{-g}  \left(
-\frac{1}{2}\nabla_{\mu}\psi\nabla^{\mu}\psi-\frac{\xi}{2}R\psi^2-U(\psi)\right),
\end{eqnarray}
and a power-Maxwell term coupled to the scalar through the function $\epsilon(\psi)$
\begin{eqnarray}\label{eq:M-S}
S_{M} &=&\int {d}^5 x\sqrt{-g} \mathcal{L}_{M},\nonumber\\
 &=&-\frac{1}{4}\int{d}^5x\sqrt{-g}
\,\epsilon (\psi) \left(F_{\mu \nu} F^{\mu \nu}\right)^{q},
\end{eqnarray} 
where $F_{\mu \nu}$ is the field strength tensor, and $q$ is a nonzero rational number with an odd denominator.
As was explained in the introduction, following the ideas from the 4DSGB theories, in the four-dimensional background, the Einstein-Hilbert Gauss Bonnet density (\ref{EHGB}) is supplemented with (\ref{eq:action-EHLambda}) together with \textcolor{black}{$\mathcal{L}_{4DSGB}$} from  (\ref{eq:GB-reg}), where now
$S_{\psi}, S_{M}$ are written for a four-dimensional spacetime. In this scenario, eq. (\ref{eq:gen-action}) takes the form  $\mathcal{L}_{g}=\mathcal{L}_{EH}$ and \textcolor{black}{ $\mathcal{L}_{\tiny{matter}}=\mathcal{L}_{4DSGB}+ \mathcal{L}_{\psi}+ \mathcal{L}_{M}$}, resulting in the following solution:

\begin{eqnarray}
ds^2&=&-f(r) dt^2+\frac{dr^2}{f(r)}+r^2 (dx^2+dy^2),\label{metric-sol}\\
f(r)&=&r^2(1-\zeta \psi(r)^{2}),\label{metric-function-sol} \\
\Lambda &=& -\dfrac{3}{2},\quad \tilde{\alpha} = \dfrac{1}{2}. \nonumber
\end{eqnarray}
The scalar fields, $\psi(r)$ and $\phi(r)$, and the Maxwell strength $F_{\mu \nu}$ are given by
\begin{eqnarray}\label{eq:psi-Frt}
\psi(r)&=&(ar-b)^{\frac{2 \xi}{4\xi-1}}, \qquad \phi(r)=\ln(r),\\
F_{rt}&=&(A_{t})'=\frac{\left[\frac{4(\zeta-\xi) \xi \zeta}{(1-4\xi)^2 q a^{\frac{4q}{2q-1}} (-2)^q}\right]^{\frac{1}{2q}}}{\left(r^{2} \epsilon(\psi)\right)^{\frac{1}{2q-1}}},
\end{eqnarray}
{where in order to have an AdS asymptotically configuration, we need to consider
\begin{equation}\label{eq:xi}
 0<\xi<\frac{1}{4}.
\end{equation}
Here, $a$} is a positive integration constant and $b$ is a positive parameter in the coupling function $\epsilon(\psi)$ as well as in the potential $U(\psi)$, which read
\begin{eqnarray}\label{eq:epsilon-U}
\epsilon(\psi)^{\frac{1}{2q-1}}&=&\frac{{\psi}^{-\frac{1}{\xi}} \left( {\psi}^{{\frac {4\,\xi-1}{2\xi}}}+b
 \right) ^{{\frac {1-6q}{2q-1}}}}{\left(20\xi-3\right) \psi^{\frac {4\xi-1}{2\xi}}+b
 \left( 4\xi+1 \right) }
, \\
U(\psi)&=&\frac{1}{(1-4\xi)^2}\,\Big(\beta_1 \psi^{2}+\beta_2 \psi^{\frac{1}{2\xi}}+\beta_3\psi^{\frac{1-2\xi}{\xi}} \nonumber\\
&+&\beta_4 \psi^4+\beta_5 \psi^{\frac{4\xi+1}{2\xi}}+\beta_6 \psi^{\frac{1}{\xi}}\Big)\label{eq:potxi}.
\end{eqnarray}
In order to be as clear as possible, the $\beta_{i}$'s  from (\ref{eq:potxi}) are reported in the Appendix \ref{Pot}.

{To} compute the $\eta/s$ ratio, the entropy $\mathcal{S}$ of this black hole is required, that can be obtained following the Euclidean approach \cite{Gibbons:1976ue,Regge:1974zd}. In this case, we have 
\begin{eqnarray}
{\cal{S}}&=&\frac{2 \pi r_h^{2} (\zeta-\xi) \Omega_{2}}{\zeta}, \label{eq:ent}
\end{eqnarray}
where, as before, $r_h$ is the location of the event horizon,  and the density entropy takes the form $s=\mathcal{S}/\Omega_{2}$, with $\zeta >\xi>0$. In order to achieve an AdS asymptotic configuration and a positive density entropy $s$, it is necessary to restrict the values of the non-minimal coupling parameter $\xi$ and the constant $\zeta$ to certain ranges, as illustrated in Figure \ref{fig3}, to ensure completeness.

\begin{figure}[ht]
\begin{center}
\includegraphics[scale=0.23]{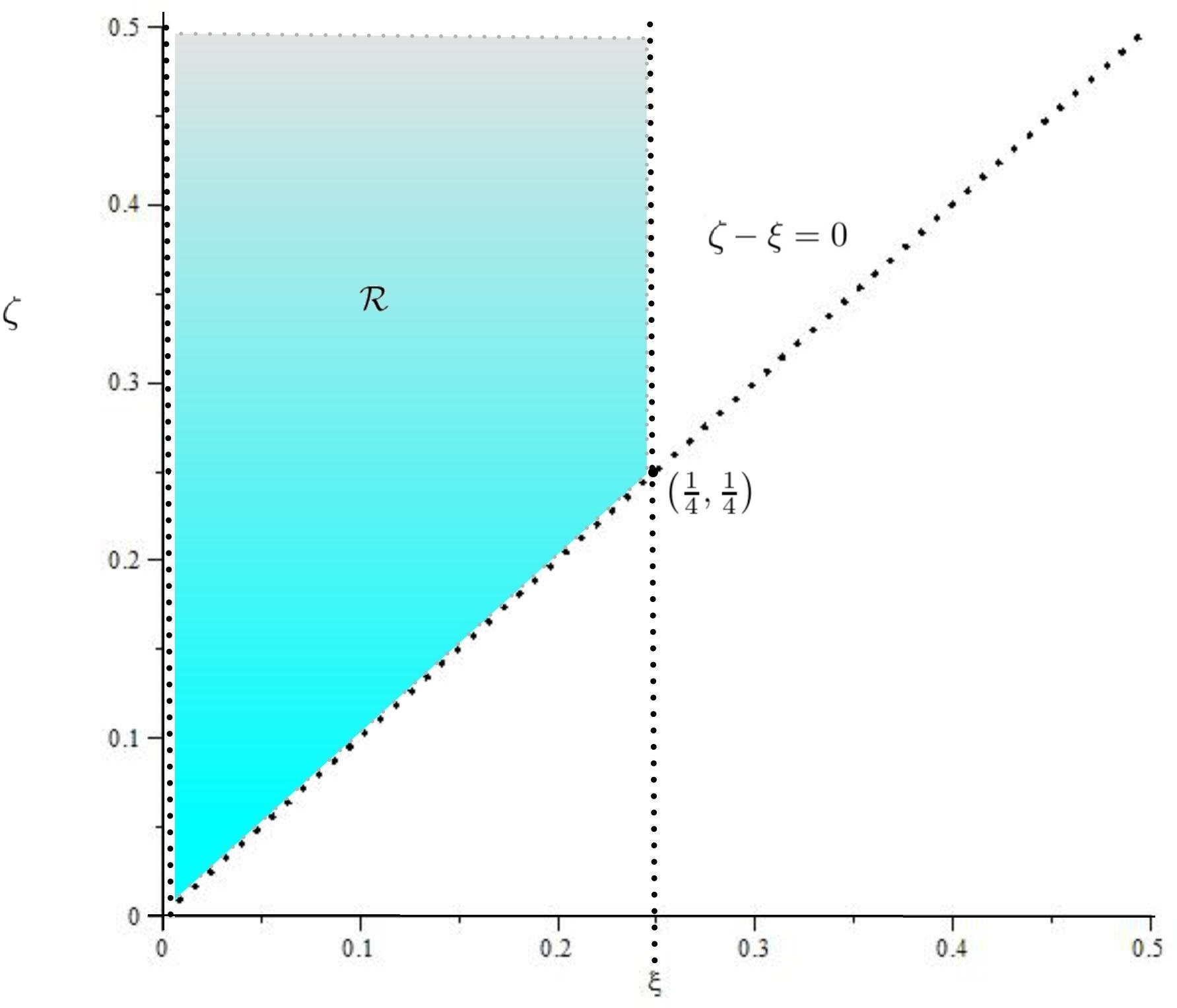}
\caption{\textcolor{black}{Graphic representation of the ranges of values for the non-minimal coupling parameter $\xi$ and the constant $\zeta$, in order to have an  asymptotically AdS configuration as well as a positive entropy $\mathcal{S}$, represented in the region $\mathcal{R}$.}}
\label{fig3}
\end{center}
\end{figure}
{Using the prescription from Sec. \ref{Section-gen-comp-shear}, we obtain
\begin{eqnarray*}
\sqrt{-g}Q^{r x} &=& \dfrac{1}{2}r^2 f(r)(h_{xy})' \Big( (1-\xi \psi(r)^2) \\ && \hspace{1cm} + 2 \tilde{\alpha} ( {f} \phi(r)'^2 - \phi'(r)f'(r)) \Big),
\end{eqnarray*}
where the $(x,y)$-component of the linear Einstein equations is given by $(\sqrt{-g}Q^{r x})'=0$. At the location of the event horizon, we note that
$$\psi^2(r)_{r=r_h}=\frac{1}{\zeta},\quad \phi'(r)\big{|}_{r=r_h}=\frac{1}{r_h}, \quad f'(r)\Big{|}_{r=r_h}=4 \pi T,$$
and  $T$ is the Hawking temperature for this configuration, which reads 
\begin{eqnarray}\label{eq:T-C}
T=\frac{(1+b \zeta ^{\frac{4\xi-1}{4\xi}}) r_h \xi}{(1-4\xi) \pi}.
\end{eqnarray}
Thus, the viscosity/density entropy ratio takes the form
\begin{equation}\label{eq:vis-C}
\dfrac{\eta}{s} = \dfrac{1}{4\pi}\left[ 1 - \dfrac{4 \xi\zeta(1+b \zeta ^{\frac{4\xi-1}{4\xi}}) }{(\zeta - \xi)(1-4\xi)} \right].
\end{equation}
Here, eq. (\ref{eq:vis-C}) allows for some remarks to be made. First of all, as the previous cases, the $\eta/s$ ratio is independent of the event horizon $r_h$, just like the examples in the introduction (see Refs. \cite{Kovtun:2004de}-\cite{Bravo-Gaete:2020lzs}). Along with the aforementioned, from (\ref{eq:vis-C}) a contribution dependent on the constants $\zeta$, $\xi$, and $b$ also emerges, in addition to the factor $1/(4 \pi)$. In fact, considering (\ref{eq:xi}) and  $\zeta>\xi$, represented in Fig. \ref{fig3}, together with $b>0$ we have
$$\dfrac{4 \xi\zeta(1+b \zeta ^{\frac{4\xi-1}{4\xi}}) }{(\zeta - \xi)(1-4\xi)}>0,$$
implying that 
$\frac{\eta}{s}<\frac{1}{4 \pi},$
and in order to obtain a non-negative ratio, we need
\begin{equation}\label{eq:G}
\mathcal{G}=1-\dfrac{4 \xi\zeta(1+b \zeta ^{\frac{4\xi-1}{4\xi}}) }{(\zeta - \xi)(1-4\xi)}\geq 0,
\end{equation}
represented via Fig. \ref{fig4}.
}

\begin{figure}[ht]
\begin{center}
\includegraphics[scale=0.15]{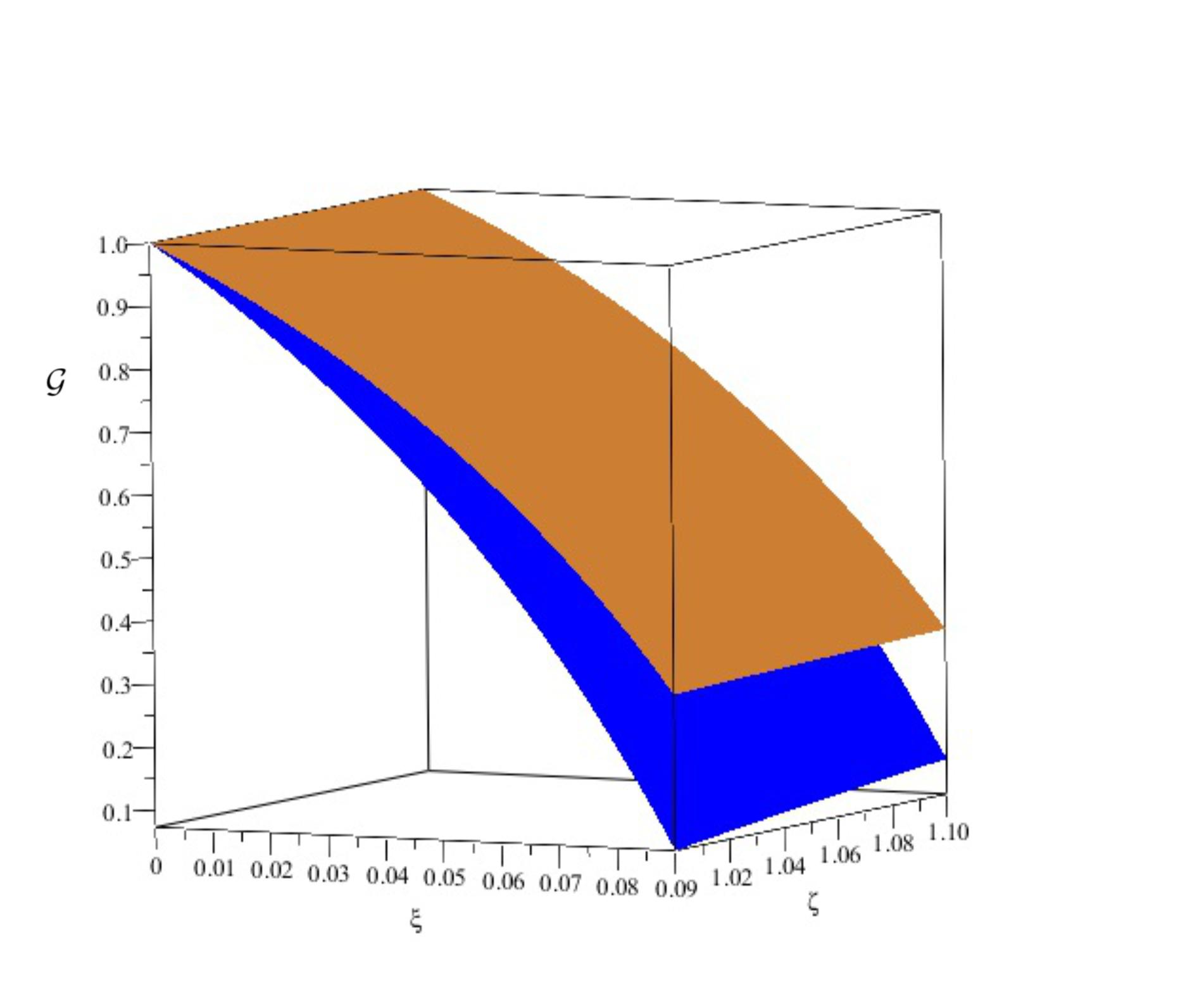}
\caption{\textcolor{black}{Graphic representation of $\mathcal{G}$ as a function of the constants $\xi$ and $\zeta$. Here, the black (gold) surface correspond to eq. (\ref{eq:G}) with $b=0.1$  ($b=0.5$) and $(\xi,\zeta) \in ]0,0.09] \times [1,1.1] \subset\mathcal{R}$, where the region $\mathcal{R}$ was showed previously in Fig. \ref{fig3}.}}
\label{fig4}
\end{center}
\end{figure}

%%%%%%%%%%%%%%%%%%%%%%%
\section{Conclusions and discussions} \label{Section-conclusions}
%%%%%%%%%%%%%%%%%%%%%%%

In this work, we have explored the well-known ratio between the shear viscosity and the entropy density, motivated by some black hole solutions with vanishing entropy. In both cases, a well-suited matter source permits to circumvent of the null-entropy situation, allowing in both cases the discussion on $\eta$, and its implications on the KSS-bound (\ref{KSS}).

Regarding the black hole solutions, our first case of study (Case A) corresponds to the one proposed in \cite{Bravo-Gaete:2021hza}, which is characterized by non-linear electrodynamics in the $(\mathcal{H}, P)$-formalism. This approach results in charged configurations of four-dimensional AdS black holes with a planar base manifold in the framework of {CG} with non-null thermodynamic quantities, in particular the entropy. Together with the above, under this scenario, it is important to note that the $\eta/s$ ratio remains constant, and is independent of any additional parameters introduced in the theory, obtaining the saturated situation from (\ref{KSS}).

{With this result at hand}, one might be interested in working with CG in particular extensions of the duality, e.g., non-relativistic physics and condensed matter systems. In those with anisotropic scaling symmetries, also known as the Lifshitz symmetry \cite{Kachru:2008yh}
\begin{equation}\label{eq:symmetry_lifshitz}
t\to\tilde{\lambda}^z\,t,\quad r\to\frac{r}{\tilde{\lambda}}, \quad {x}\to\tilde{\lambda}
{x},  \quad {y}\to\tilde{\lambda}
{y},
\end{equation}
 it is possible to find in the literature black holes configurations, dubbed as Lifshitz black holes. Here, $t$, $r$,${x}$ and $y$ are the time, radial and spatial coordinates respectively, $z$ is known as the dynamical exponent, responsible for the anisotropic scaling between $t$, ${x}$ and $y$, while $\tilde{\lambda}$ is a non null constant. As was mentioned earlier, although the equations of motion are of the fourth order, CG is renormalizable and ghost-free. In ref. \cite{Bravo-Gaete:2021kgt} it was shown that dilatonic fields, this is
\begin{eqnarray}
\mathcal{L}_{\tiny{matter}}&=&-\frac{1}{2} \nabla_{\mu} \phi \nabla^{\mu} \phi-\frac{1}{4} e^{\lambda \phi} F_{\mu \nu} F^{\mu \nu},\label{eq:dil-scalar}\nonumber
\end{eqnarray}
with $F_{\mu \nu}$ defined previously in (\ref{eq:Fmunu}), can support Lifshitz black holes in four dimensional CG, with or without a non-minimal scalar field $\psi$ (eq. (\ref{eq:psi-action})). In both examples, it is straightforward to check that the bound (\ref{KSS}) remains saturated despite the introductions of the dynamical exponent $z$ and the non minimal parameter $\xi$. {\color{black}This establish a remarkable difference with other models with quadratic contributions (see for example \cite{Brigante:2007nu}). The reasons behind this behavior are not clear, and it} would be interesting to analyze the shear viscosity in other models containing CG, to provide a physical meaning of the saturation.

In contrast to Case A, our second case of study (Case B) involves a four-dimensional reduction of the Einstein-Gauss-Bonnet action coupled with a cosmological constant (\ref{EHGB}). The matter source is characterized by a non-minimally coupled scalar field $\psi$ (\ref{eq:psi-action}) and a power-Maxwell term coupled to the scalar through the function $\epsilon(\psi)$ (\ref{eq:M-S}). We show that the non-minimal coupling parameter $\xi$, as well as the constants $\zeta$ and $b$, have an impact on the viscosity $\eta$. As a result, the $\eta/s$ ratio (\ref{KSS}) is also affected. {The
dependence of the ratio in terms of the nonminimal parameter continues to exist when the vacuum solution is supplemented with a non-linear source in the Pleb\'anki formalism \cite{Bravo-Gaete:2022mnr}.} This opens up new avenues for future research, such as a more in-depth analysis of the dimensional extension of these cases or the exploration of other models beyond GR.

It is important to note that the focus of the work is not on whether the KSS bound (\ref{KSS}) is violated or fulfilled. As we mentioned in the introduction, there are various examples, including our study, that demonstrate that the inequality (\ref{KSS}) is not always valid. Instead, our aim is to study the impact of the different parameters from the gravity theories, given by the cases A and B, on the viscosity $\eta$ and the $\eta/s$ ratio (\ref{KSS}) in two different black hole solutions, enriching the ranges of possibilities. On the other hand, and for the sake of completeness, the second approach to calculate $\eta$ involves the Kubo formula \cite{Kovtun:2004de, Policastro:2001yc, Son:2002sd, Kovtun:2003wp}, and details are presented in Appendix \ref{Kubo} . {As expected}, both methods yield identical expressions for the $\eta/s$ ratio.

\begin{acknowledgments}

The authors would like to thank Mar\'ia Montserrat Ju\'arez-Aubry and 
Daniel Highita for useful discussion and comments on this work. M.B. is supported by PROYECTO INTERNO UCM-IN-22204, L\'INEA REGULAR.
\end{acknowledgments}

%%%%%%%%%%%%%%%%%%%%%%%%%%%%%%%%%%%%%%%%%%%%%%%%%%%%%%%%%%%%%%%%%%%%%
\section{Appendix } \label{Section-appendix}
%%%%%%%%%%%%%%%%%%%%%%%%%%%%%%%%%%%%%%%%%%%%%%%%%%%%%%%%%%%%%%%%%%%%%

%%%%%%%%%%%%%%%%%%%%%%%%%%%%%%%%%%%%%%%%%%%%%%%%%%
\subsection{Coefficients $\beta_{i}$'s present in the potential $U(\psi)$ from Eq. (\ref{eq:potxi})}\label{Pot} 
%%%%%%%%%%%%%%%%%%%%%%%%%%%%%%%%%%%%%%%%%%%%%%%%%%

In this subsection, we present the coupling constants $\beta_{i}$'s from the potential $U(\psi)$ 
present in the eq. (\ref{eq:potxi}), and are given by
\begin{eqnarray*}
\beta_{1}&=&\frac{(12\xi-2)
(16\xi -3)\xi}{2}, \qquad \beta_{2}={4(12\xi-2)b\xi^2},\nonumber\\ \nonumber\\
\beta_{3}&=&2 \xi^2 b^2,\nonumber\\ \nonumber\\
\beta_{4}&=&\frac{\zeta}{4 q} \left( 20 \xi  -3 \right)  \big[ 
 (\zeta-2\xi )  ( 16 \xi -2) q
\nonumber\\
&-&4 \xi (\zeta-\xi)  \big]
,\nonumber\\ \nonumber\\
\beta_{5}&=&\frac {\xi b\zeta}{q} \big\{\left( 24-128q\right) {\xi}^{2}+ \big[  \left( 16+64
\zeta \right) q\nonumber\\
&-&24\zeta -2 \big] \xi-2\zeta
 \left( 4q-1 \right) 
\big\}
 , \nonumber\\ \nonumber\\
\beta_{6}&=&\frac { \left( 4\xi+1 \right) \xi {b}^{2} \left[\zeta \left( 2q-1 \right)  -\xi \left( 4
q-1 \right)  \right] \zeta}{q}.
\end{eqnarray*}

%%%%%%%%%%%%%%%%%%%%%%%%%%%%%%%%%%%%%%%%%%%%%%%%%%%%%%%
\subsection{Shear viscosity via the Kubo formula}\label{Kubo}
%%%%%%%%%%%%%%%%%%%%%%%%%%%%%%%%%%%%%%%%%%%%%%%%%%%%%%%

To provide a comprehensive analysis, it is worth noting that the shear viscosity for these cases can be obtained by using the related Kubo formula {\cite{Kovtun:2004de, Policastro:2001yc, Son:2002sd, Kovtun:2003wp}}. To carry out this procedure, we will activate appropriate metric perturbations in the bulk and thus obtain the response function. For this, we start with the perturbations $\delta^{(1)}g^{\mu \nu}=-h^{\mu\nu}$, $\delta^{(2)}g^{\mu \nu}=h^{\mu\alpha}h^{\nu}_{\alpha}$, $\delta^{(1)}\sqrt{-g}=0$, $\delta^{(2)}\sqrt{-g}=-\sqrt{-g}h_{\mu\nu}h^{\mu\nu}/4$, $\delta^{(1)}R_{\mu \nu}$, $\delta^{(2)}R_{\mu \nu}$, via the general action (\ref{eq:gen-action}). Here,  $\delta^{(1)}R_{\mu \nu}$ and $\delta^{(2)}R_{\mu \nu}$ are given by
\begin{eqnarray*}
&&\delta^{(1)}R_{ij}=\partial_{\mu}(\delta^{(1)}\Gamma^{\mu}_{ij})-\partial_{i}(\delta^{(1)}\Gamma^{\mu}_{j\mu})+(\delta^{(1)}\Gamma^{\mu}_{\mu\rho})\Gamma^{\rho}_{ij}\nonumber\\
&&+\Gamma^{\mu}_{\mu\rho}(\delta^{(1)}\Gamma^{\rho}_{ij})-(\delta^{(1)}\Gamma^{\mu}_{i\rho})\Gamma^{\rho}_{\mu j}-\Gamma^{\mu}_{i\rho}(\delta^{(1)}\Gamma^{\rho}_{\mu j}),\\
&&\delta^{(1)}\Gamma^{k}_{ij}=\frac{1}{2}(\partial_{i}h^{k}_{j}+\partial_{j}h^{k}_{i}-\partial^{k}h_{ij}),\\
&&\delta^{(2)}R_{\nu\alpha}=\partial_{\mu}(\delta^{(2)}\Gamma^{\mu}_{\nu\alpha})-\partial_{\nu}(\delta^{(2)}\Gamma^{\mu}_{\mu\alpha})+(\delta^{(1)}\Gamma^{\mu}_{\mu\rho})\delta^{(1)}\Gamma^{\rho}_{\nu\alpha}\nonumber\\
&&-\delta^{(1)}\Gamma^{\mu}_{\nu\rho}\delta^{(1)}\Gamma^{\rho}_{\mu\alpha}+\delta^{(2)}\Gamma^{\mu}_{\mu\rho}\Gamma^{\rho}_{\nu\alpha}+\Gamma^{\mu}_{\mu\rho}\delta^{(2)}\Gamma^{\rho}_{\nu\alpha}\nonumber\\
&&-\delta^{(2)}\Gamma^{\mu}_{\nu\rho}\Gamma^{\rho}_{\mu\alpha}-\Gamma^{\mu}_{\nu\rho}\delta^{(2)}\Gamma^{\rho}_{\mu\alpha},\\
&&\delta^{(2)}\Gamma^{k}_{ij}=-\frac{h^{kl}}{2}(\partial_{i}h_{jl}+\partial_{j}h_{il}-\partial_{l}h_{ij}).
\end{eqnarray*}
Considering the first-order perturbations $\delta^{(1)}g_{\mu\nu}=h_{\mu\nu}$, we can write the transverse and traceless (TT) tensor perturbation in a general way to the shear viscosity, where $\partial_{\alpha}h_{\mu\nu}=0$ and $h\equiv\eta^{\mu\nu}h_{\mu\nu}=0$, and considering the metric perturbation (\ref{eq:t-metric}) with $h(r)=f(r)$, we have
\begin{eqnarray}
\delta^{(1)}R_{xy}&=&-\frac{1}{2}r^{2}\Box\Psi-(2f+rf^{'})\Psi,\nonumber\\
&=&-\frac{1}{2}r^{2}\Box\Psi+R^{(0)}_{xx}\Psi,
\end{eqnarray}
where $R^{(0)}_{xx}$ denotes any of the (diagonal) components of the zeroth-order Ricci tensor of the
background metric. Now, the first-order perturbation of the Einstein tensor takes the form
\begin{eqnarray}
\delta^{(1)}G_{xy}&=&-\frac{1}{2}r^{2}\Box\Psi-R^{(0)}_{xx}\Psi-\frac{1}{2}R^{(0)}h_{xy},\nonumber\\
&=&-\frac{1}{2}r^{2}\Box\Psi+G^{(0)}_{xx}\Psi,\nonumber\\
&=&-\frac{1}{2}r^{2}\Box\Psi+\frac{1}{2}T^{(0)}_{xx}\Psi,
\end{eqnarray}
{where the second order for $\delta^{(2)}G_{xy}$ is given by}
\begin{eqnarray}
\delta^{(2)}G_{xy}&=&\frac{1}{4}\delta^{(1)}G_{xy}+\frac{1}{8}\Psi[(2f+rf^{'})\Psi^{'}+rf\Psi^{''}],\nonumber\\
                  &=&\frac{1}{4}\delta^{(1)}G_{xy}+\frac{1}{8}\Psi\left(R^{(0)}_{xx}\Psi^{'}+rf\Psi^{''}\right).
\end{eqnarray}
Combining all elements with Kubo’s formula relates the components of the viscosity to the two-point function of corresponding components of the stress-tensor $T_{xy}$, given by 
\begin{eqnarray}
\eta=\Im(\langle\,T_{xy}(\vec{k}_{1},\omega)T_{xy}(\vec{k}_{2},\omega)\rangle^{'})|_{\vec{k}_{1},\vec{k}_{1}\to0,\omega\to0}.\label{imgreen}
\end{eqnarray}
Here, $\vec{k}$ is the spatial momentum, $\omega$ is the frequency, and the prime subscript means that the overall energy-momentum conserving delta function has been removed. From the AdS/CFT correspondence,  we have that
\begin{eqnarray}
\langle\,T_{xy}(\vec{k}_{1},\omega)T_{xy}(\vec{k}_{2},\omega)\rangle=\frac{\delta^{(2)}S}{\delta\,h_{xy}(\vec{k}_{1})\delta\,h_{xy}(\vec{k}_{2})},
\end{eqnarray}
where $S$ represent the general action (\ref{eq:gen-action}). Finally, we have that the shear viscosity takes the form for:
\begin{eqnarray}
&&\textbf{Case A:}\, \eta=\frac{1}{4\pi}{2 \pi} r_h^{2}\left(\frac{3 \zeta \alpha_1-2 \alpha_2}{3 \zeta^{2}}\right)=\frac{1}{4 \pi} s,\nonumber\\
&&\textbf{Case B:}\, \eta=\frac{1}{4\pi} s \left[ 1 - \dfrac{4 \xi\zeta(1+b \zeta ^{\frac{4\xi-1}{4\xi}}) }{(\zeta - \xi)(1-4\xi)} \right],
\end{eqnarray}
where for our computations, through the two techniques the results converge.

%%%%%%%%%%%%%%%%%%%%%%%%%%%

\end{document}